\documentclass[%
twocolumn,
 amsmath,amssymb,
 aps,
 longbibliography,
 footinbib,
]{revtex4-2}

\usepackage{physics}
\usepackage{ascmac}
\usepackage{graphicx}
\usepackage{dcolumn}
\usepackage{bm}
\usepackage{here}
\usepackage{wrapfig}
\usepackage{color}
\usepackage[dvipsnames]{xcolor}
\usepackage{CJKutf8}
\usepackage{hyperref}

\hypersetup{
    colorlinks,
    linkcolor={red!50!black},
    citecolor={blue!50!black},
    urlcolor={blue!80!black}
}

\begin{document}
\begin{CJK}{UTF8}{min}

\title{Quantum soliton-trains of strongly correlated impurities in Bose--Einstein condensates}

\author{Hoshu Hiyane}
\email{hoshu.hiyane@oist.jp}
\affiliation{Quantum Systems Unit, Okinawa Institute of Science and Technology Graduate University, Onna, Okinawa 904-0495, Japan}

\author{Thomas Busch}
\affiliation{Quantum Systems Unit, Okinawa Institute of Science and Technology Graduate University, Onna, Okinawa 904-0495, Japan}

\author{Thom\'as Fogarty}
\affiliation{Quantum Systems Unit, Okinawa Institute of Science and Technology Graduate University, Onna, Okinawa 904-0495, Japan}

\begin{abstract}
    Strongly correlated impurities immersed in a Bose--Einstein condensate (BEC) can form a periodic structure of tightly localized single atoms due to competing inter- and intra-species interactions, leading to a self-organized pinned state.
    In this work, we show numerically that the impurities in the self-pinned state form a soliton-train, as a consequence of a BEC-mediated attractive self-interaction and ordering due to the exclusion principle. 
    The dynamics of the impurities possess the characteristics of bright matter-wave soliton-trains as often seen in classical fields; however, in the few impurities cases, the detailed nature of collisions is determined by their quantum statistics.
\end{abstract} 

\maketitle
\end{CJK}



Solitons, and more generally nondispersive solitary waves, appear in many physical systems. The historical and paradigmatic example are shallow water waves, whose dynamics can be modelled by the Korteweg--De Vries equation, which is well-known to possess soliton solutions~\cite{book_Lamb_1980}.
Over the last couple of decades dark and bright solitons have been extensively studied in nonlinear optical systems~\cite{Zakharov_1972,Emplit_1987,Krokel_1988,Kivshar_1989,Kivshar_1993,Malomed_2005} and in atomic Bose--Einstein condensates (BECs)~\cite{Burger_1999,Busch:00,Carr_2000_1,Carr_2000_2,Denschlag_2000,Khayakovich_2002,Strecker_2002,Marchant_2013,Kartashov_2019}, which are both systems that are well described by classical fields whose dynamics is governed by the nonlinear Schr\"odinger equation (NLSE). 
As atomic BECs are amenable to clean and highly controllable experiments many different realisations of solitons have been explored, ranging from scalar dark~\cite{Burger_1999,Busch:00,Denschlag_2000} and bright~\cite{Carr_2000_2,Strecker_2002} solitons to vector dark-dark~\cite{Hoefer_2011,Yan_2012} and dark-bright~\cite{Busch_2001,Nistazakis_2008,Becker_2008,Hamner_2011} solitons. 
More recently, complex soliton structures in two-dimensional two-component BECs, called Townes solitons, have been experimentally observed~\cite{hassani_2021}.

Beyond single solitons, highly excited soliton arrays, known as soliton-trains, have also been theoretically and experimentally investigated. 
While the exact solutions of the free-space NLSE contains such states for both repulsive (dark)~\cite{Carr_2000_1} and attractive (bright)~\cite{Carr_2000_2} interaction, their experimental realization is not easy. While more controlled ways to create bright solitons have been suggested \cite{Edmonds:18}, bright solitons are usually created by first preparing a stable condensate with repulsive interactions and then suddenly quenching the scattering length from positive to negative. However, this excites a modulational instability that results in the cloud breaking up into bright soliton-trains~\cite{Strecker_2002,Carr_2004}. 
This process is therefore inherently uncontrollable and results in a collection of bright solitons of differing widths and particle numbers, making single solitons and soliton-train states hard to study systematically.
In nonlinear optics, on the other hand, the controllable generation of temporal soliton-trains is possible using ultrashort laser pulses with high repetition rates~\cite{Mollenauer_1990,Mamyshev_1991,Haus_1996,Kivshar_1998,Buryak_2002,Chen_2012}. 
The ability to deterministically create and control matter-wave soliton trains could have important applications in quantum engineering, such as in high-precision sensing with atom interferometry~\cite{Ramanathan_2011,Martin_2012,Polo_2013,Wales_2020,Szigeti_2020}.

Recently it has been suggested that strongly repulsive bosonic impurities immersed in a BEC can localize and self-organize into a periodic atomic array due to the competition between impurity-impurity and impurity-BEC interactions~\cite{Keller_2022}. 
In this work, we show that such self-localized impurities form soliton-train states in the regime where their velocity is below the speed of sound of the BEC. 
Such bright soliton-trains can be understood as the spatial matter-wave counterpart of temporal optical solitons in nonlinear media, with the nonlinear coupling mediated by the BEC matter-wave. 
However, contrary to the similarities with optics, here only few atoms are sufficient for these nonlinearities to appear~\cite{Bruderer_2008,Keller_2022} and the solitonic impurities can be treated fully quantum mechanically. The system therefore supports quantum soliton-train states, and we show that quantum properties, such as the statistics of the impurity atoms (bosonic or fermionic), have a strong influence on their nonequillibrium properties.


In the following we first briefly review the system of two coupled one-dimensional quantum gases at ultralow temperatures 
where the first component is a weakly correlated atomic BEC described in the mean-field limit, while the second one is a minority component of $N\lll N_{\text{BEC}}$ bosonic impurities \cite{Keller_2022}.
The impurities are assumed to strongly interact with each other so that their many-body dynamics can be described using the Tonks--Girardeau (TG) gas model~\cite{Girardeau:60}.
The inter-component interaction is described by a repulsive density-density coupling of strength $\gamma$, which we scale relative to the strength of the mean-field interaction of the condensate $g$. We assume equal masses $m$ for both components with the dynamics of the coupled system described by
\begin{align}
    i\dot\Psi=&\left[-\frac{1}{2}\pdv[2]{x}+V_{\rm BEC}+|\Psi|^2+\gamma\rho\right]\Psi,
    \label{eq:GPE}\\
    i\dot\phi_n=&\bqty{-\frac{1}{2}\pdv[2]{x}+V_{\rm TG}
    +\gamma|\Psi|^2}\phi_n.
    \label{eq:SPSE}
\end{align}
Here $\Psi(x,t)$ is the condensate order parameter obeying mean-field Gross--Pitaevskii equation (GPE)~\cite{pitaevskii_stringari2016} and $\phi_n(x,t)$ are the single particle states needed to describe the TG gas using the Bose--Fermi mapping theorem~\cite{Girardeau:60}. All quantities are scaled with respect to the characteristic length scale $x_0=\hbar^2/(mg)$, which is related to the condensate-healing length, and time scale $t_0=mg^2/\hbar^3$, so that all wavefunctions are in units of $\sqrt x_0$. 
Since the density of the TG gas, $\rho(x,t)$, is equivalent to the density of a gas of spin-polarised fermions, at zero temperature it can simply be written as
$\rho(x,t)=\sum^N_{n=1}|\phi_n(x,t)|^2$.
We assume the BEC to be trapped in a ring potential of length $L_{\rm BEC}$ so that it has a flat density profile, while the TG gas is confined in a smaller box trap $V_{\rm TG}(x)$ of length $L< L_{\rm BEC}$~\cite{Keller_2022}. 
This setup allows us to explore the motion of the impurities without considering edge effects from the condensate.

\begin{figure}
    \centering
    \includegraphics[width=\linewidth]
    {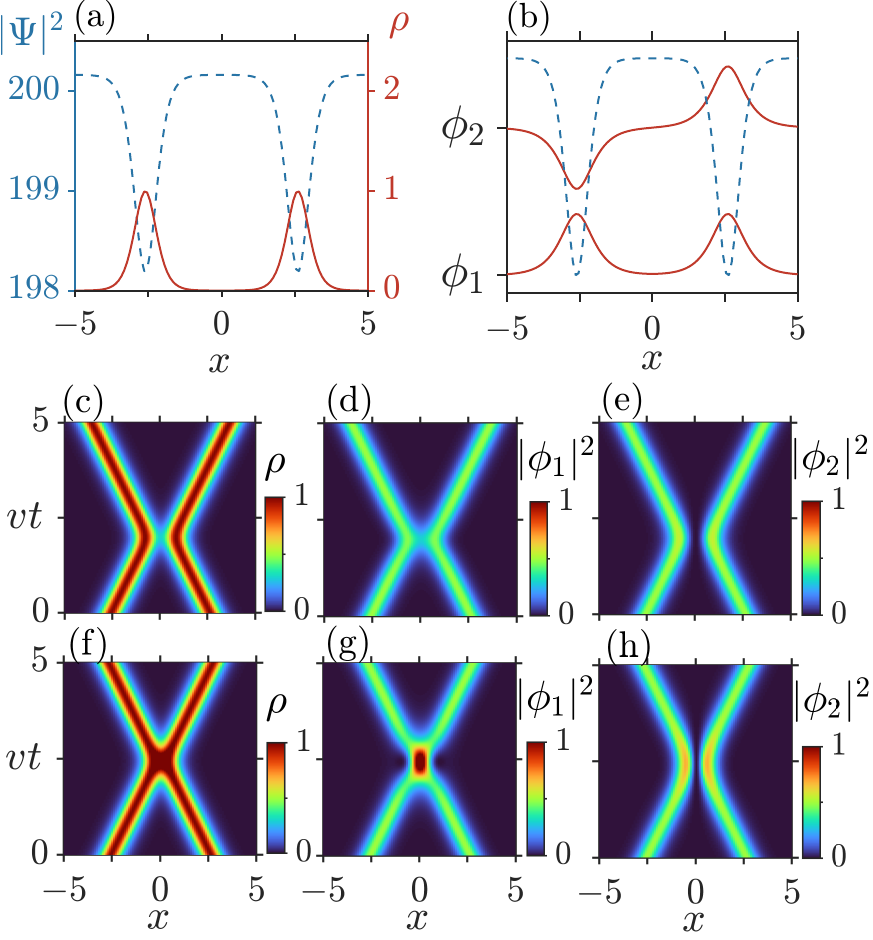}
    \caption{
    (a) Density profiles and (b) single particle eigenstates of the ground state of the impurities (red solid lines) pinned in a BEC with density profile given by the blue dashed line at $t=0$.
    Evolution of the density of the impurities, $\rho(x,t)$,
    after receiving a momentum kick of (c) $v=0.15$ and (f) $v=1$. 
    Evolution of the respective single particle states $|\phi_1(x,t)|^2$ and $|\phi_2(x,t)|^2$ for (d-e) $v=0.15$ and (g-h) $v=1$.
    Parameters are $N_{\rm BEC}=5000$, $L_{\rm BEC}=25$, $L=15$, and $\gamma=2$.
    }
    \label{fig:N2_dynamics}
\end{figure}

The ground state of the coupled system can be found by solving the coupled equations in a self-consistent manner by means of imaginary time evolution for Eq.~\eqref{eq:GPE} and exact diagonalization for Eq.~\eqref{eq:SPSE}. 
When the systems are decoupled for $\gamma=0$, the TG gas is delocalized in the box potential and may be considered as a quasi-superfluid state.
For sufficiently strong coupling strength $\gamma$ the impurity atoms localize individually in the mean-field potential provided by the BEC. The many-body groundstate then becomes a regularly spaced array of pinned impurities (see Fig.~\ref{fig:N2_dynamics} (a)).
This array of impurities therefore represents a \textit{self-pinned} insulator state within a matter-wave lattice,
and the distance between impurities is set by the Fermi momentum $k_F=N\pi/L$~\cite{Miyakawa_2004,Keller_2022}.

In this self-pinned state, an effective model for the impurity eigenstates and their dynamics can be constructed by taking the Thomas--Fermi approximation for the BEC component.
The solution to Eq.~\eqref{eq:GPE} is then given by~\cite{Stringari:96} 
\begin{align}
	\Psi(x,t)\approx\sqrt{\mu_{\rm TF}-\gamma\rho(x,t)}
         e^{i\beta t}
 ,
	\label{TFA}
\end{align}
where $\beta$ is a real-valued, constant phase and the equilibrium chemical potential is
$
	\mu_{\rm TF}=\mu_0\qty(1+\gamma N/N_{\rm BEC}),
$
with $\mu_0=N_{\rm BEC}/L_{\rm BEC}$ being the chemical potential evaluated at $\gamma=0$~\cite{Keller_2022}.
Under this approximation the Schr\"odinger equation for the impurities Eq.~\eqref{eq:SPSE} can be rewritten as
\begin{equation}
	i\dot\phi_n(x,t)=\bqty{-\frac{1}{2}\pdv[2]{x}-\gamma^2\rho(x,t)}\phi_n(x,t),\label{TFASE}
\end{equation}
with the impurities coupled to one another through the effective attraction term $-\gamma^2 \rho(x,t)$ and the eigenstates being orthogonal to each other $\int dx\,\phi_n(x,t)\phi_m(x,t)=\delta_{nm}$. 
Equation~\eqref{TFASE} resides in the large group of $N$-coupled NLSEs ($N$-CNLSEs), which are also often referred to as Manakov equations~\cite{Manakov_1973}. They possess (bright) soliton-trains as stationary solutions that are thoroughly investigated in the field of nonlinear optics~\cite{Gordon_1983,Radhakrishnan_1997,Kanna_2001,Kanna_2003,Kanna_2006}. 
In particular, Eq.~\eqref{TFASE} simplifies for a single impurity ($n=N=1$) with density $\rho=|\phi_1|^2$ which results in the NLSE with self-attraction and the solution is the celebrated scalar bright soliton, $\rho(x)=(\gamma/2)^2\sech^2(\gamma^2x/2)$~
\footnote{
Since the stationary self-pinned solutions of Eqs.~\eqref{eq:GPE} and~\eqref{eq:SPSE} show that no overlap between neighbouring impurities exists (see Fig.~\ref{fig:N2_dynamics}~(a)), one can treat them separately and it is therefore possible to describe the ground state properties by using effective single impurity NLSEs with attractive self-interaction~\cite{Keller_2022}
}.

\begin{figure}
    \centering
    \includegraphics[width=\linewidth]{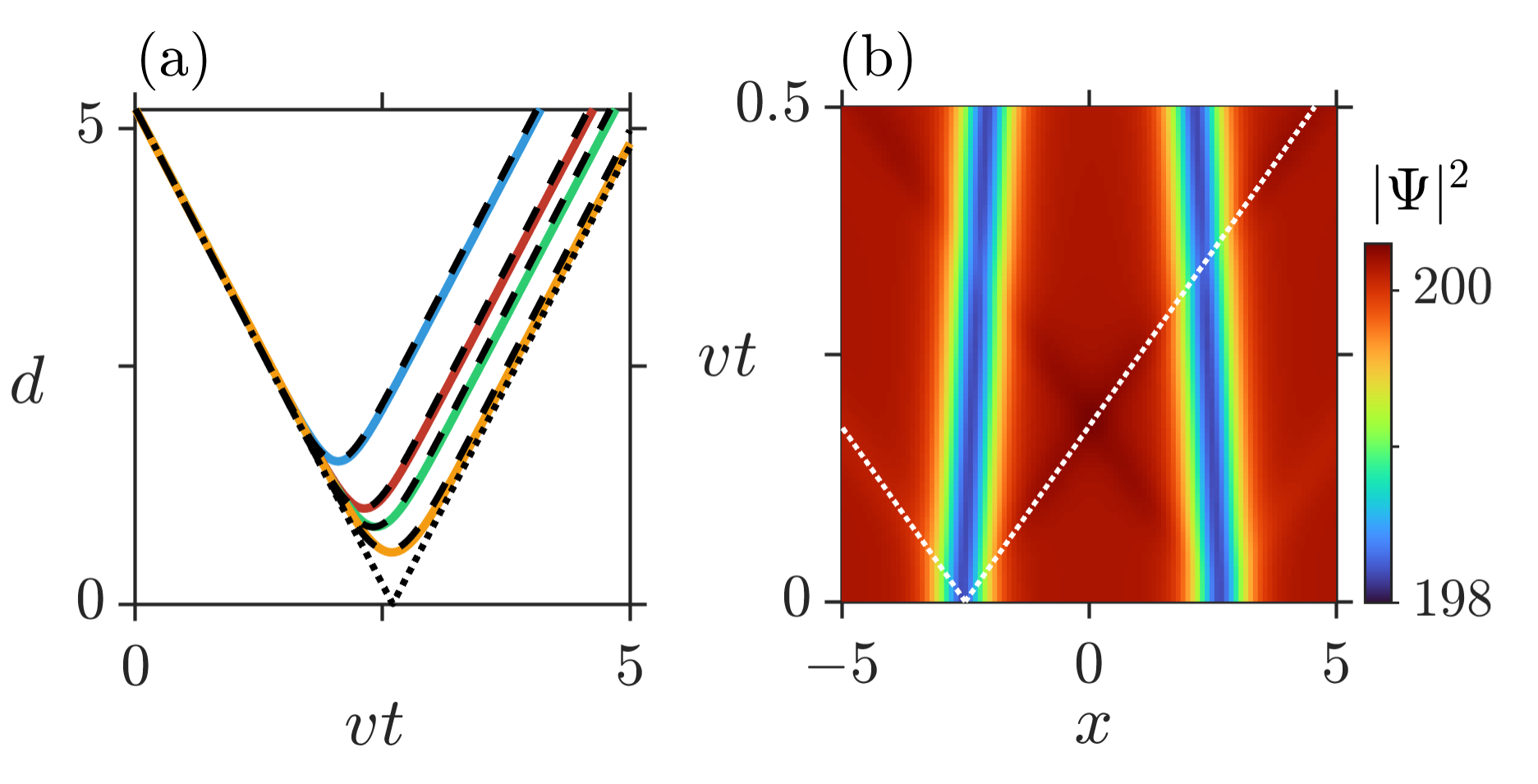}
    \caption{
    (a) Average distance between impurities from Eq.~\eqref{eq:distance}, for $v=0.2$ (blue), $0.6$ (red), $1$ (green), $3$ (yellow) calculated by solving the full system in Eqs.~\eqref{eq:GPE} and~\eqref{eq:SPSE}. Dashed lines are calculated from the effective model Eq.~\eqref{TFASE}.
    Dotted line is the corresponding quantity for two noninteracting classical particles: $d(t)=|d(0)-2vt|$.
    (b) BEC density evolution after an impurity momentum kick of $v=1$. White dotted line corresponds to $x=\pm v_{\rm B}t-d(0)/2$, where $v_{\rm B}$ is the Bogoliubov speed of sound for a homogeneous BEC.
    }
    \label{fig:phonon_avgdistance}
\end{figure}

To go beyond this effective model and explore the nonequilibrium properties of impurities in the BEC, it is necessary to solve the full Eqs.~\eqref{eq:GPE} and~\eqref{eq:SPSE} numerically.
However, in the following we will show that in certain regimes it is possible to describe the behavior observed by just considering the soliton-train solutions that the $N$-CNLSE \eqref{TFASE} allows for. 
For this we will confirm that the wave profile is maintained during the free evolution, and that the shape and phase difference are maintained after a collision \cite{book_Lamb_1980,Martin_2007,Martin_2008,malomed_2014}.
Our first step is therefore to solve Eqs.~\eqref{eq:GPE} and~\eqref{eq:SPSE} for the minimal system of $N=2$ impurities and to study the collisional properties by applying opposite momentum kicks $e^{\pm i v x}$
to the initially localised parts of the wavefunction in Fig.~\ref{fig:N2_dynamics}~(a) (see Appendix.~\ref{appendix:initialization} for the detailed numerical technique).
The resulting density evolutions of the impurities for two different relative initial velocities is shown in  Figs.~\ref{fig:N2_dynamics} (c) and (f).
One can immediately see that in both cases the impurities stay localized and maintain their individual shape after the kick and also after the collision process. However, while for the collision with large relative velocity, $v=1$, the two localised solitonic impurities strongly overlap (and possibly cross), for the collision with lower velocity, $v=0.15$, a clear repulsion between the impurities is visible. 
To investigate the impurity scattering we calculate the relative distance 
\begin{equation}
    d(t)=\int dx_1dx_2|x_1-x_2||\Phi(x_1,x_2,t)|^2\;,
    \label{eq:distance}
\end{equation}
which is shown for different velocities in Fig.~\ref{fig:phonon_avgdistance} (a). 
One can see that with increasing relative velocities the impurities approach and overlap with each other more and more, and their trajectories approach that of noninteracting classical particles represented by the dotted line.
We also show that the results obtained from the effective description Eq.~\eqref{TFASE} (dashed lines) agree quite well with the result from the full coupled system of Eq.~\eqref{eq:GPE} and Eq.~\eqref{eq:SPSE}. This shows that we remain in the regime of validity of the Thomas--Fermi approximation as the impurity velocity never exceeds the Landau critical velocity, which can be estimated by the Bogoliubov speed of sound for a homogeneous BEC to be $v_{\rm B}\approx \sqrt{N_{\rm BEC}/L_{\rm BEC}}\sim14$. We note, however, that immediately after the instantaneous momentum kick, a weak burst of phononic excitations with velocity $\pm v_{\rm B}$ appears in the BEC (see Fig.~\ref{fig:phonon_avgdistance} (b))~\cite{Andrews_1997}, nevertheless they do not alter the dynamics of the impurities as their amplitude is small.
The effective $N$-CNLSE \eqref{TFASE}
therefore well describes the system and its dynamics in the regime where phononic excitations in the BEC can be neglected, which in turn means that the soliton-train solutions provide an accurate description of the dynamics of the $N$ impurities~\footnote{It is important to note that even though these soliton-trains are embedded into the background of a BEC, they are inherently different to the well-studied dark-bright soliton, e.g.~\ Refs.~\cite{Busch_2001,Mistakidis_2019}, since the density inhomogeneities in the BEC are small and do not carry the phase defects that define dark solitons.
They are also different to the Bose--Fermi composite solitons investigated previously in Refs.~\cite{Karpiuk_2004,Salerno_2005,Karpiuk_2006,Santhanam_2006,Adhikari_2007,Rakshit_2019_NJP,Rakshit_2019_scipost,Desalvo_2019}, where strong inter-species attraction surpasses the intra-species repulsion among bosons resulting in the effective attraction in the BEC leading to a formation of a bright soliton in the BEC supported by fermions.}.

This agreement then allows us to understand the velocity dependence of the scattering process due to the overlapping phenomena of the solitons~\cite{Aitchison_1991,Malomed_1991,Stegeman_1999,Hedge_2022} from the fact that these solitonic impurities can be seen as composite vector-solitons formed by the single particle states $\phi_1$ and $\phi_2$ of Eq.~\eqref{TFASE}, which are coupled by their total density $\rho$.
While the scattering behaviour of scalar many-body Bose-condensed bright solitons is purely determined by their relative phase~\cite{book_Lamb_1980,Martin_2007,Martin_2008,malomed_2014}, for the soliton-trains formed by the single particle states $\phi_1$ and $\phi_2$, one has to carefully examine the eigenstates' individual dynamics. 
For this one can see in Fig.~\ref{fig:N2_dynamics} (b) that the phase of state $\phi_1$ is flat, whereas the phase for state $\phi_2$ possess a $\pi$-phase jump at $x=0$.
This jump leads to an effective repulsion between the localized parts on the left and the right hand side of $\phi_2$, which will dominate over the kinetic energy at small distances. 
For low relative velocities the impurity density therefore never shows a crossing~\cite{Aitchison_1991,Malomed_1991,Stegeman_1999,Hedge_2022}, as the coupling between $\phi_1$ and $\phi_2$ ensures that the state $\phi_1$ follows the trajectory given by $\phi_2$ as can be seen from Fig.~\ref{fig:N2_dynamics} (d) and (e). This results in the repulsive collision dynamics observed in Fig.~\ref{fig:N2_dynamics} (c).
On the other hand, for large relative velocities the kinetic energy can be high enough to lead to a closer approach for $\phi_2$ and therefore a significant overlap in the density can become indistinguishable from a crossing (see Figs.~\ref{fig:N2_dynamics} (f-h)). 
We emphasize that the dynamics shows composite (vector) bright soliton-trains formed by orthogonal fermionic states, whose properties are strongly influenced by exclusion statistics; therefore, the dynamics of the $N$-CNLSE \eqref{TFASE} (or the original coupled  Eqs.~\eqref{eq:GPE} and \eqref{eq:SPSE}) cannot be described by that of the single NLSE, leading to distinctively different bright soliton dynamics compared to simple BEC systems.

\begin{figure}
    \centering
    \includegraphics[width=\linewidth]{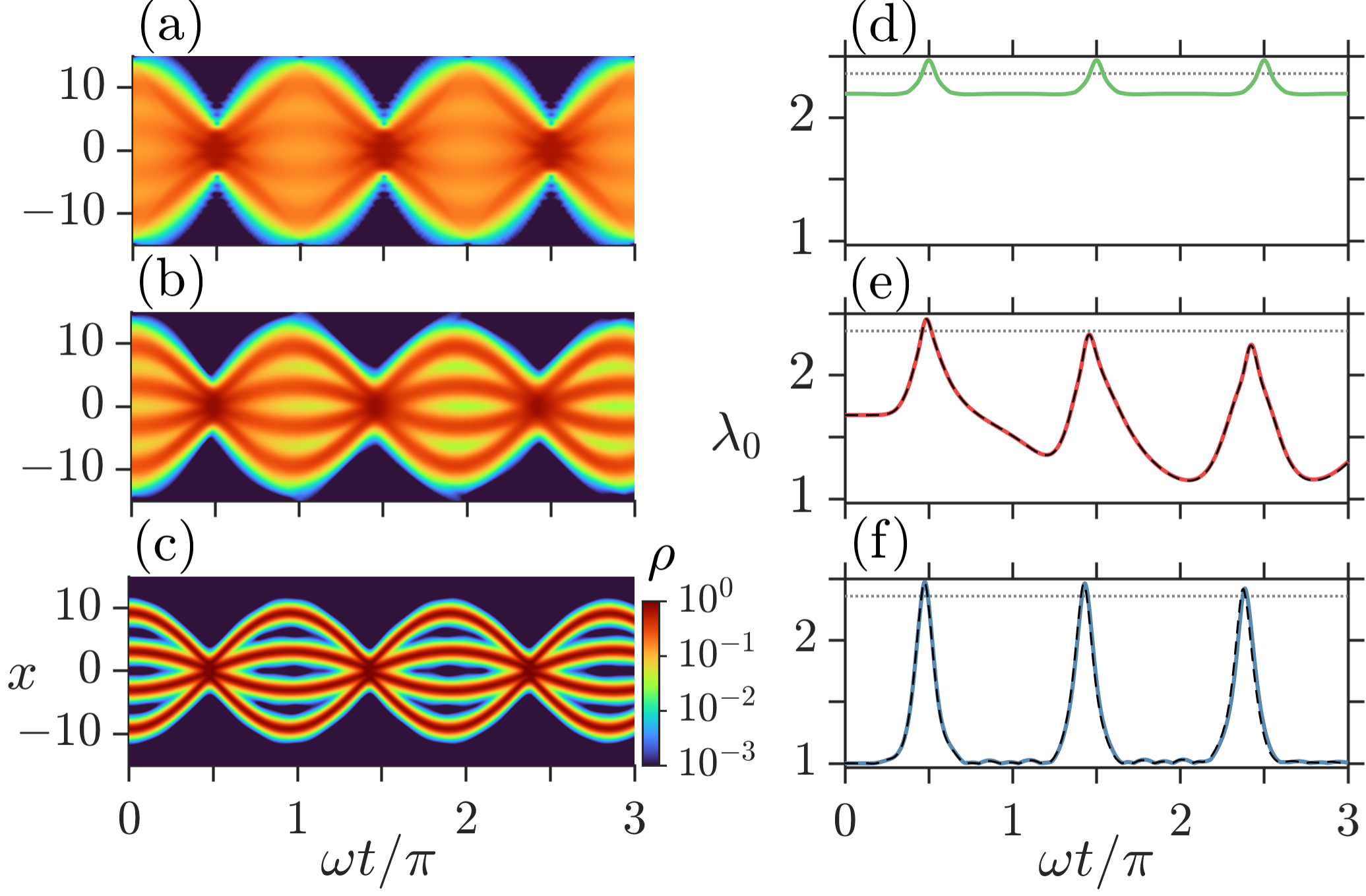}
    \caption{
    (left column) Density evolution of four impurities coupled to a BEC after a sudden quench from an external box potential to a harmonic trap with trapping frequency $\omega=0.15$.
    (right column) Largest occupation number, $\lambda_0(t)$, of the RSPDM.
    Solid line is obtained by evolving the coupled equations~\eqref{eq:GPE} and \eqref{eq:SPSE}, while the dashed lines in panels (e) and (f) are obtained by evolving Eq.~\eqref{TFASE}.
    Dotted line shows the coherence $\lambda_0$ of the ground state of the TG gas in the harmonic trap with frequency $\omega$. Coupling strengths are $\gamma=0$, $\gamma=1$, and $\gamma=1.8$ from top to bottom and the 
    other parameter choices are $N_{\rm BEC}=10^4$, $L_{\rm BEC}=50$, and $L=30$.
    }
    \label{fig:density_lam0}
\end{figure}


To explore the dynamical behaviour of larger trains of impurity solitons we quench the system by instantaneously changing the external trap from a box to a harmonic oscillator potential, $V_{\rm TG}(x)=\omega^2 x^2/2$, with $\omega$ in units of $1/t_0$.
The resulting density dynamics for $N=4$ impurities is shown in Fig.~\ref{fig:density_lam0}~(a-c) for different coupling strengths: for $\gamma=0$ the impurities are not coupled to the BEC and undergo free evolution, whereas for $\gamma=1.8$ the system is deep in the pinned regime where the impurities behave as soliton-trains. In between ($\gamma=1$) the system is in an intermediate regime where the impurities are quasi-localized but still have finite overlap with one another.
In all cases, the time evolution of the density profile shows the harmonic trap-induced breathing mode with periodic collisions and revivals of the impurities every $t\sim \pi/\omega$, which is a realization of the quantum Newton's cradle \cite{kinoshita_2006,Caux_2019} in a coupled two component system.
Furthermore, one can note that in the case of $\gamma=1.8$, the impurities are localized even after all four impurities have collided multiple times providing further evidence of being the soliton-train. 

Since the evolution Eqs.~\eqref{eq:GPE} and \eqref{eq:SPSE} can be used to simulate both the bosonic TG impurities as well as free fermionic impurities, it is interesting to explore how their dynamics are affected by differences in their quantum statistics. However, since the Bose--Fermi mapping theorem ensures that the density evolution is identical for both exchange symmetries, $\rho_{\rm F}(x,t)=\rho_{\rm TG}(x,t)$, it can not be used to discern any differences between them \cite{Girardeau:60}. 
On the other hand, the reduced single particle density matrices (RSPDMs) and the (experimentally measurable) momentum distributions can be different for bosonic or fermionic impurities. The RSPDM of the impurities is given by $\varrho(x,x',t)=\int \Phi^*(x,x_2,\dots,x_N,t)\Phi(x',x_2,\dots,x_N,t) dx_2,\dots,dx_N$, where $\Phi$ is the respective many-body wavefunction (see Appendix~\ref{appendix:RSPDM} for the detailed discussion).
This matrix can be diagonalised to find the so-called natural orbital basis, $\varrho(x,x',t)=\sum_n\lambda_n(t) \varphi^*_n(x,t)\varphi_n(x',t)$, where $\lambda_n(t)$ are the occupation probabilities of the orbitals $\varphi_n(x,t)$. The largest occupation number $\lambda_0(t)$ characterizes the degree of coherence in the system and takes a maximal value of $\lambda_0\approx N$ for a weakly interacting coherent BEC, and is $\lambda_0=1$ for noninteracting fermions.

The coherence within the TG system strongly depends on the coupling strength $\gamma$ (see Figs.~\ref{fig:density_lam0} (d-f)).
For $\gamma=0$ at time $t=0$, the TG impurities overlap and strongly interact with each other, which reduces the coherence compared to weakly interacting bosons, but still allows for a large degree of coherence on the order of $\lambda_0\sim\sqrt{N}$~\cite{girardeau_2001}.
Such a system may still be considered as a quasi-superfluid.
When the particles approach each other around the collision time, $t=\pi/2\omega$, the coherence increases due to quasi-condensation~\cite{Rigol_2004}.
Since for $\gamma=0$ the impurity system is integrable, the collisions and the increase in coherence happen periodically. 
For $\gamma=1.8$ the pinned TG state is initially fully incoherent due to the isolation of particles from each other. 
However, during collisions, the coherence temporarily increases rapidly as the impurities overlap, reaching about the same value as in the uncoupled situation.
The quench dynamics in the solitonic regime therefore allows to realize a dynamical transition between different phases, with the periodic oscillations between the pinned insulator and the quasi-superfluid state being supported by the coupling to the BEC.
Finally, in the intermediate regime of $\gamma=1$, the impurities overlap throughout the dynamics, and their coherence can be seen to decay after repeated collisions. 
This can be attributed to the fact that the BEC-mediated coupling among the TG atoms breaks the integrability of the system.
We note that in all cases the maximum velocity of the impurities is around $v\sim L\omega/2=2.25$, which is much smaller than the Landau critical velocity estimated above. 
Thus, although we do not show it here, the amplitudes of phonons excited in the BEC are negligible, and the Thomas--Fermi approximation also holds in this case. The dynamics observed from the effective nonlinear equation (dashed lines in Figs.~\ref{fig:density_lam0} (e-f)) are therefore indistinguishable from the dynamics described by the full system. 

\begin{figure}
    \centering
    \includegraphics[width=\linewidth]{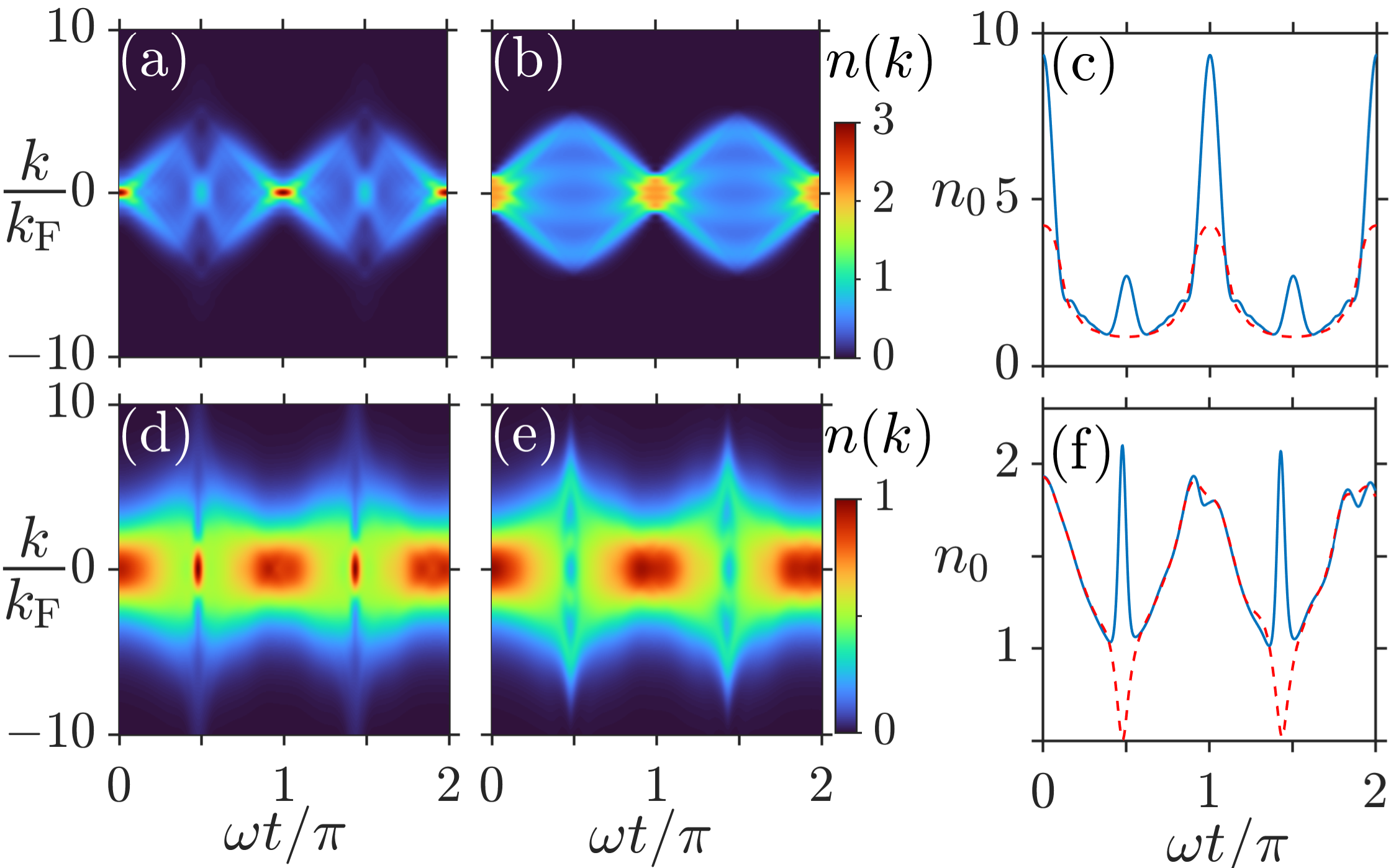}
    \caption{(top row) Momentum distribution $n(k,t)$ following a trap quench in the free regime $\gamma=0$ for (a) TG and (b) Fermi impurities.
    (c) Peak of the momentum distribution at $k=0$ for TG (solid line) and Fermi (dashed line) impurities. 
    (bottom row) Equivalent dynamics in the self-pinning regime ($\gamma=1.8$). The parameter choice is the same as in Fig.~\ref{fig:density_lam0}.
    }
    \label{fig:nkt}
\end{figure}


For a fermionic system the coherence is unchanged throughout the dynamics, $\lambda_n(t)=1$ for $0\leq n\leq N-1$, due to the separable mean-field ansatz between the noninteracting fermions and the condensate.
However, the dynamics of the fermionic momentum distribution $n(k,t)=\int e^{-ik(x-x')} \varrho(x,x',t)dx dx'$ can be compared to the bosonic TG one, and distinct characteristics due to different particle symmetries can be seen. The upper row of Fig.~\ref{fig:nkt} shows the momentum distributions for free evolution, $\gamma=0$,  and one can see notable differences. In particular, the TG impurities show a significant peak at $k=0$ whenever the atoms collide (see panel (c)). 
This is due to the impenetrable hardcore character of their scattering, which reverses the impurity momenta after each collision~\cite{Atas_2017,Atas_2017_2}.
For fermions, this peak is absent as they simply pass through each other.  The lower row in Fig.~\ref{fig:nkt} shows the same quantities, but now for the situation where the impurities strongly interact with the BEC, $\gamma=1.8$. One can see that the momentum distribution of the TG and fermionic impurities are equivalent at all times between collisions, highlighting the incoherent single particle nature of both systems when deep in the pinned regime where we have $\lambda_n(t)=1$. 
However, during the collisions the different scattering mechanisms are enhanced by the BEC-mediated effective attraction. 
The $k=0$ mode for the fermions is drastically depleted as the attractive interactions accelerate the particles through one another when their densities overlap.
Conversely, the $k=0$ mode for TG impurities is increased as the particles scatter with larger kinetic energy.
Solitonic impurities in the BEC therefore act in different ways according to their quantum statistics whenever the particles collide and overlap, while at all other times they have an incoherent single particle nature.

In conclusion, we have studied the dynamics of strongly correlated impurities, namely bosons in the TG regime or fermions, which are coupled to a BEC in the mean-field limit.
Investigating the collision dynamics of two impurities in the pinned limit, we provide strong evidence that the impurities satisfy the characteristic scattering properties of soliton trains when the velocity of the impurities is small enough such that the amplitude of the generated phonons in the BEC are negligible.
The system can then be accurately described by a set of $N$-coupled nonlinear Schr\"odinger equations, which allows one to identify the origin of the formation of the impurity brght soliton-train as a consequence of exclusion principles and BEC-mediated self-attraction.
Unlike the soliton states formed by a (fully coherent) classical field such as in BECs or nonlinear optical systems, the few-body nature of the impurity soliton-train allows to study correlations among impurities and the quantum statistical difference between TG bosons and noninteracting fermions.
Indeed, 
the coupling to the BEC can stabilize the impurities against dispersion, allowing the localized pinned state to survive even after repeated collisions and
the induced breathing mode exhibiting dynamical transitions between insulating and quasi-superfluid states.
Moreover, TG bosons can be distinguished from spinless fermions in the momentum distribution whenever the impurities overlap, showing distinct collisional effects enhanced by the presence of the BEC. 
It is also known that the $N$-CNLSE \eqref{TFASE} can show interesting intensity redistributions (or energy exchanges), leading to inelastic scattering among the solitons~\cite{Kanna_2001,Kanna_2003}. 
To study this inelastic scattering along with the quantum statistical properties of solitonic impurities would be an intriguing future work.
Another important extension would be the inclusion of correlations between the BEC and impurities, allowing to explore beyond mean-field effects on both the impurity localization and their soliton dynamics. 

\acknowledgements
This work was supported by the Okinawa Institute of
Science and Technology Graduate University. The authors are grateful for the Scientific Computing and
Data Analysis (SCDA) section of the Research Support
Division at OIST.
The authors thank Tetsuro Nikuni and Tim Keller for insightful discussions.
T.F. acknowledges support from JSPS KAKENHI Grant No. JP23K03290. T.F. and T.B. are also supported by JST Grant No. JPMJPF2221.

\appendix

\section{Initialization and dynamics\label{appendix:initialization}
}

The order parameter of the Bose gas, $\Psi$, and the fermionic single particle wavefunctions, $\phi_n$, obeying Eq.~(1) and (2) in the main text, are coupled via the density of the respective other species,  $\rho=\sum^N_{n=1}|\phi_n|^2$ and $|\Psi|^2$.
Therefore, the two equations have to be solved self-consistently to find the ground state.
We perform this computation by finding the ground state and the density of the mixture using imaginary time evolution with a split-step Fourier transform technique for the BEC~\cite{Bao_2003}, and an exact diagonalization scheme for the fermionic single particle states.
In the first instance, the ground state and the density of the mixture at $\gamma=0$ can be found analytically as it is the solution of the homogeneous BEC and fermionic particles in a box.
This ground state is then used as a trial wavefunction for the bosonic component to obtain the BEC ground state with a small but finite coupling strength $\gamma>0$ by evolving Eq.~(1) in imaginary time.
The obtained density of the BEC is then inserted into the single particle Hamiltonian Eq.~(2) diagonalized to find the updated density of the impurities. This procedure is iterated until convergence is reached.
The newly obtained ground state can be again used as an initial trial state to find the ground state with a slightly larger coupling strength $\gamma$ and this procedure is repeated for small increases in the coupling until the final groundstate is found for the desired coupling strength~\cite{Salerno_2005}.

Using the obtained ground state as an initial state, we compute the time evolution for $N=2$ as presented in Figs.~\ref{fig:N2_dynamics} and \ref{fig:phonon_avgdistance} and for $N=4$ as presented in Figs.~\ref{fig:density_lam0} and \ref{fig:nkt}.
In each case, the dynamics is initiated by quenching impurity component as follows.
For $N=2$, a momentum kick with strength $v$ is performed ($\psi_n(x,t=0^+)=\psi_n(x,t=0^-)e^{-ivx}$ for $x<0$ and $\psi_n(x,t=0^+)=\psi_n(x,t=0^-)e^{ivx}$ for $x>0$) in order to monitor the time-evolution in free space.
For $N=4$, an external harmonic trap is added {\it only} for the impurities to study the collision of four solitonic impurities.
The subsequent dynamics of the mixture are again simulated using the split-step Fourier transform method~\cite{Bao_2003}.


\section{Reduced single particle density matrix of Tonks--Girardeau gas
\label{appendix:RSPDM}
}

The equation of motion for the TG gas couples to the BEC in the form of a single particle potential, which means that the Hamiltonian reads
\begin{align}
    \hat H=\sum^N_{n=1}\pqty{-\frac{1}{2}\pdv[2]{x_n}+\gamma |\Psi(x_n)|^2}+g\sum^N_{1\leq i<j \leq N}\delta(x_j-x_k),
\end{align}
with the TG limit taking $g\to \infty$.
The infinite repulsive interaction prevents two particles from occupying the same position which implies a constraint on the many-body wavefunction
\begin{align}
    \Psi_{\rm TG}(x_1,\cdots, x_i,\cdots, x_j,\cdots, x_N)=0
    \quad
    \text{for}
    \quad
    x_i=x_j\;.
    \label{eq:TGcondition}
\end{align}
This constraint allows to map the TG gas to spin-polarized fermions described by the single particle Hamiltonian
\begin{align}
    \hat H_{\rm sp}=-\frac{1}{2}\pdv[2]{x}+\gamma |\Psi(x)|^2\;,
    \label{eq:singleparticleH}
\end{align}
with eigenfunctions $\phi_n(x)$.
The fermionic many-body wavefunction can be constructed by the Slater determinant $\Psi_{\rm F}=\det[\phi_i(x_j)]_{1\leq i,j\leq N}$ with its bosonic counterpart found through appropriate symmetrization via $\Psi_{\rm TG}=\prod_{k>j}\text{sgn}(x_k-x_j)\Psi_{\rm F}$.
This is the famous Bose--Fermi mapping theorem~\cite{Girardeau:60}, which allows to simplify the computation of the physical quantities of the TG gas dramatically as the main computational effort is to obtain the fermionic single particle states $\phi_n(x)$.

In the main text, making use of the Bose--Fermi mapping, the largest eigenvalue (coherence) of the reduced single particle density matrix (RSPDM) of the TG gas and the momentum distribution is computed.
The RSPDM is defined as $\varrho(x,x',t)=\expval{\hat \psi^\dagger(x,t)\hat \psi(x',t)}$ and its Fourier transform leads to the momentum distribution $n(k,t)=(2\pi)^{-1}\int dxdx'e^{ik(x-x')}\varrho(x,x',t)$~\cite{girardeau_2001}.
In the case of the noninteracting fermions, the RSPDM takes the form of
\begin{align}
    \varrho(x,x',t)=\sum^N_{n=1}\phi^*_n(x,t)\phi_n(x',t),
    \label{eq:fermiRSPDM}
\end{align}
and therefore, the momentum distribution is a simple sum of the Fourier transformed single particle states: $n(k,t)=\sum^N_{n=1}|\phi_n(k,t)|^2$, where $\phi_n(k,t)=(\sqrt{2\pi})^{-1}\int dxe^{ikx}\phi_n(x,t)$.
As shown by Pezer and Buljan~\cite{Pezer_2007}, there exists an efficient computational method to compute the RSPDM of the TG gas, which can be written analogously to the fermionic counterpart as
\begin{align}
    \varrho(x,x',t)=\sum^N_{n=1}\sum^N_{m=1}\phi^*_n(x,t)A_{nm}(x,x',t)\phi_m(x',t),
    \label{eq:pezerbuljanRSPDM}
\end{align}
where $N$ by $N$ matrix $\hat A(x,x',t) = A_{nm}(x,x',t)$ is defined as 
\begin{align}
    \hat A(x,x',t)
    =&(\hat P(x,x',t))^{-1}\det\hat P(x,x',t),
    \label{eq:pezerbuljanA}
    \\
    [\hat P(x,x',t)]_{nm}=&\delta_{nm}-2\int^{x'}_{x}\phi^*_n(y,t)\phi_m(y,t)dy.
    \label{eq:pezerbuljanP}
\end{align}
We compute the RSPDM using Eqs.~\eqref{eq:pezerbuljanRSPDM}--\eqref{eq:pezerbuljanP} for the TG gas and Eq.~\eqref{eq:fermiRSPDM} for the fermions, which yields the momentum distribution via the Fourier transform presented in Fig.~\ref{fig:nkt} in the main text.

The eigenstates and eigenvalues of the RSPDM are called natural orbitals $\tilde\phi(x,t)$ and occupation numbers $\lambda_n(t)$
\begin{align}
    \varrho(x,x',t)=\sum_{n=0}\lambda_n(t)
    \tilde\phi^*_n(x,t)\tilde\phi_n(x',t),
\end{align}
which can give further information about the many-body properties.
For instance, the RSPDM of non-interacting fermions at zero-temperature, Eq.~\eqref{eq:fermiRSPDM}, is diagonal with natural orbital being exactly the same as the single particle states $\tilde\phi_n(x,t)=\phi_n(x,t)$ with unit occupation number $\lambda_{0\leq n< N}=1 $ and $\lambda_{n\geq N}=0$ indicating its incoherent nature.
A fully coherent BEC, on the other hand, possesses only one orbital, $\lambda_n=N\delta_{n,0}$ and $\tilde\phi_0(x,t)$ becomes the Bose order parameter.
The largest eigenvalue, $\lambda_0$, is also known to be a measure of the coherence and is used to characterize the coherent dynamics of the TG-impurities in a BEC, which is presented in the main text in the right column of Fig.~\ref{fig:density_lam0}.


\bibliography{ModifiedManuscript}

\end{document}